\documentstyle[12pt]{article}
\def\beq{\begin{equation}}
\def\eeq{\end{equation}}
\def\gord{$ \raisebox{-.3ex}{$\stackrel{>}{_{\sim}}$} $}

\begin{document}
\begin{titlepage}
\begin{center}
{\Large \bf Theoretical Physics Institute \\
University of Minnesota \\}  \end{center}
\vspace{0.15in}
\begin{flushright}
TPI-MINN-92/64-T \\
November 1992
\end{flushright}
\vspace{0.2in}
\begin{center}
{\Large \bf On the Current Correlators in QCD at Finite Temperature \\ }
\vspace{0.4in}
{\bf V.L. Eletsky$^{\dagger}$,} \\
Theoretical Physics Institute, University of Minnesota  \\
Minneapolis, MN 55455 \\
\vspace{0.2in}
{\bf B.L. Ioffe} \\
Institute of Theoretical and
Experimental Physics, Moscow 117259, Russia \\
\vspace{0.5in}
{\bf   Abstract  \\ }
\end{center}

Current correlators in QCD at a finite temperature $T$ are considered
from the viewpoint of operator product expansion.
It is stressed that at low $T$ the heat bath must be represented by
hadronic, and not quark-gluon states. A possibility to express the results in
terms of $T$-dependent resonance masses is discussed. It is demonstrated
that in order $T^2$ the masses do not move and the only phenomenon which
occurs is a parity and isospin mixing.

\vskip2.0in
\hrule height .2pt width 3in
\noindent$^{\dagger}$Permanent address: Institute of Theoretical and
Experimental Physics, Moscow 117259, Russia.
\end{titlepage}

In the last years there has been an increasing interest in the study of the
current correlators in QCD at finite temperatures. The hope is that
investigating the same correlators both at high temperatures where the
state of quark-gluon plasma is expected, and at low temperatures where the
hadronic phase persists, a clear signal for a phase transition could be
found. In this aspect of special interest is the temperature dependence of
hadronic masses which manifest themselves through the behavior of correlators
at large space-like distances, the appearance of poles in the correlation
functions, etc. The calculations of the correlators are performed in the
lattice simulations as well as by various analitical methods. (For reviews
see, e.g., Refs.\cite{deg,sh}).

Among the analitical methods one of the most popular is the extension of QCD
sum rule approach to the case of finite temperature. The idea is as follows.
In the QCD sum rule method hadronic masses are obtained by investigation of
current correlators and to a large extent are determined by the
values of vacuum condensates. Therefore, if the
temperature dependence of the condensates were known, it would also allow to
find the temperature dependence of hadronic masses.

Unfortunately, in carrying out this program certain wrong steps were taken
and a misunderstanding exists in the literature. However, a clear
understanding of the possibilities of finite temperature QCD sum rules,
or more generally, the possibilities of the operator product expansion
in determination of current correlators at finite temperatures, is
essential especially in comparing the results obtained in this approach
with the ones in lattice calculations.

In this note we formulate (although partly it was done before in the
literature) the basic points of the QCD sum rule method at finite temperature
and principal results which are and could be obtained by this method.

The object under consideration is a thermal average of a current correlator
defined as

\beq
C(q,T)=\langle i\int d^4 x e^{iqx}T\{j^{+}(x),j(0)\}\rangle _T =
\frac{\sum_{n}\langle n|i\int d^4 x e^{iqx}T\{j^{+}(x),j(0)\}
e^{-H/T}|n\rangle}{\sum_{n}\langle n|e^{-H/T}|n\rangle}
\label{c}
\eeq
where $j(x)$ is a colorless current which can have Lorentz and flavor indeces
and can be a spinor, $H$ is the QCD Hamiltonian
and the sum is over all states of
the spectrum. It is assumed that $q^2$ is space-like, $q^2<0$, and $|q^2|$
is much larger than a characteristic hadronic scale, $|q^2|\gg R_{c}^{-2}$,
where $R_c$ is the confinement radius, $R_{c}^{-1}\sim 0.5\: GeV$.

We consider the case of temperatures $T$ below the phase
transition temperature $T_c$. In principle, the summation over $n$ in
eq.(\ref{c}) can be performed over any full set of states $|n\rangle$ in the
Hilbert space. It is clear however that at $T<T_c$ the suitable set of
states is the set of hadronic states, but not the quark-gluon basis.
Indeed, in this case the original particles forming
the heat bath, which is probed by
external currents, are hadrons. The summation over the quark-gluon
basis of states would require to take into account the full range of their
interaction.
This point was first mentioned in Ref.(\cite{dei}). In the previous
papers\cite{prev} devoted to the extension of QCD sum rules to finite
temperatures and even in some following papers\cite{fhs} this was not
understood and the summation over $|n\rangle$ at low $T$ was performed
in the quark-gluon basis without account of confinement.

At low $T\ll T_c$ the expansion in $T/T_c$ can be performed. The main
contribution comes from the pion states, $|n\rangle =|\pi\rangle ,\;
|2\pi\rangle ,\;...$. In the chiral limit, when $u$ and $d$ quarks and pions
are massless, the expansion parameter is $T^2/f_{\pi}^2$ where
$f_{\pi}=1333\, MeV$ is the pion decay constant: the one-pion contribution
to Eq.(\ref{c}) is proportional to $T^2/f_{\pi}^2$, two-pion contribution
is of order $T^4/f_{\pi}^2$, etc\cite{dei}. The contributions of massive
hadronic states $|n\rangle$ are suppressed by $\exp(-m_{n}/T)$.

At $|q^2|\gg R_{c}^{-2}$ the operator product expansion (OPE) of
$T\{j^{+}(x),j(0)\}$ {\em on the light cone} can be performed so the
coefficient functions are $T$-independent. Here generally is a difference
in comparison with the case of $T=0$, where expansion at small $x_{\mu}$
(or near the tip of the light cone) takes place. This follows from the fact
that at $q_0\gord |q^2/2m_{\pi}$ the matrix elements
$\langle n|T\{j^{+}(x),j(0)\}|n\rangle$ are
similar to the matrix elements of deep inelastic lepton-hadron scattering,
where the process proseeds on the light cone, $x^2\sim 1/|q^2|$, but the
longitudinal distances, along the light cone are large and do not decrease
with an increase of $|q^2|\cite{bl}$.
In the interesting special cases, when $q_0=0$ or ${\bf q}=0$  and $q_0$ pure
imaginary, an OPE near the tip of the light cone can be performed ans a few
terms in this expansion must be taken into account.
In this expansion operators with
nonzero spin $s$ appear, unlike the case $T=0$ where only spin zero operators
contribute to the vacuum average. This fact is now well understood in the
calculations done by QCD sum rule method at finite temperature\cite{eek,hkl}
as well as at finite hadronic density\cite{dl}-\cite{dr}.
At low temperatures where
the pion states dominate in the chiral limit, the matrix elements of $s\neq 0$
operators are proportional to $T^{s+2}$. It must be mentioned that even for
$s=0$ operators and their vacuum expectation values (v.e.v.) not all of the
techniques which were successful at $T=0$ can be applied at $T\neq 0$.
For example, the factorization hypothesis which works well for v.e.v. of
four-quark operators at $T=0$ cannot be directly applied at $T\neq 0$,
because pion intermediate states should be accounted for\cite{hkl,e1}.

In QCD sum rule method at $T=0$ the v.e.v. of a current correlator
calculated by OPE, is on the other hand represented by the contributions of
physical states using a dispersion relation in $q^2$ and in this way
parameters of physical states (in particular, hadronic masses) were obtained.
In this aspect the case of $T\neq 0$ dramatically differs from the case
of $T=0$. The matrix elements

\beq
\sum\ ^{'}\langle n|i\int d^4 x e^{iqx}T\{j^{+}(x),j(0)\}|n\rangle e^{-E_n/T}
\label{sum}
\eeq
where the sum is performed over all degrees of freedom of the state
$|n\rangle$, are functions of two Lorentz invariants, $q^2$ and
$\nu_{n}=p_{n}q$ (or $s_{n}=(p_{n}+q)^2$)
where $p_n$ is the momentum of the state $|n\rangle$. Therefore, to consider
analitical properties  of the amplitudes in question and represent them
in terms of physical states, it is necessary to take into account that they
have discontinuities in both $q^2$ and $s_n$ (and also in the crossing
channel, $u_{n}=(p_{n}-q)^2$). This was not done in
QCD sum rule calculations at finite temperature\cite{eek,hkl} and
density\cite{c}-\cite{hp}. (In Refs.\cite{dl,dr} an attempt was made to partly
account for these effects at finite density).

It means that in writing down the dispersion relation in $q^2$ it is necessary
to specify at which values of other invariants it takes place.
For example, if $\nu =nq=q_0$ ($n=(1,0,0,0)$ is the time-like vector
characterizing the heat bath) is fixed, then the contributions of intermediate
physical states in $s$- and $u$-channel must be taken into account.
In the case of isospin 1 vector current $j_{\mu}(x)$ it follows, e.g. that
besides the $\rho$-meson pole $\pi\rho$ states are also contributing.
It must be kept in mind that the representation of the physical spectrum
as a lowest resonance pole plus continuum which is standard in the QCD sum
rule method at $T=0$ is not suitable in the problem in question, because
at least two poles and two different continua in two ($-q^2$ and $s$)
channels should be taken into account. Further, it is evident that an
effective mass of the lowest hadronic state to be obtained
in this approach, depends on the relation between $q_0$ and
$|{\bf q}|$. E.g., it should be different for the cases
$q_0=0$, $|{\bf q}|\gg R_{c}^{-1}$ and imaginary
$q_0$, $|q_0|\gg R_{c}^{-1}$, ${\bf q}=0$.

For all these reasons the QCD sum rule approach does not generally look very
promissing for the problem of calculation of masses of lowest hadronic
states at finite temperature contrary to the case of $T=0$,
where this method proved to be very effective.
However, the calculation of $q$- and $x$-dependences (the latter was recently
advocated in\cite{sh}) for
various current correlators at finite temperature
by the OPE on the light cone is still of a considerable interest.
It would be very important,
if lattice calculations of the same correlators at intermediate $q^2$
and/or $x^2$ could also be performed, because it would then give a
possibility to determine  $T$-dependences of various condensates.

As was shown in Ref.\cite{dei}, the situation essentially simplifies
if we confine ourselves to the first order term in the expansion in $T^2$.
In this case apart from the vacuum state it is sufficient to take into
account only the one-pion state in the sum over $|n\rangle$ in Eq.(\ref{c}).
Its contribution can be calculated using PCAC and current algebra.
In Ref.\cite{dei} the correlators of isospin 1 vector and axial currents
were considered and the following relations were found

\begin{eqnarray}
C_{\mu\nu}^{V}(q,T)&=&(1-\epsilon)C_{\mu\nu}^{V}(q,0)+
\epsilon C_{\mu\nu}^{A}(q,0) \nonumber\\
C_{\mu\nu}^{A}(q,T)&=&(1-\epsilon)C_{\mu\nu}^{A}(q,0)+
\epsilon C_{\mu\nu}^{V}(q,0)
\label{mix}
\end{eqnarray}
where $C_{\mu\nu}^{V,A}(q,T)$ are the correlators of $V$ and $A$ currents at
finite temperature and $C_{\mu\nu}^{V,A}(q,0)$ are the same correlators at
$T=0$. In the chiral limit, $\epsilon =(1/3)(T^2/f_{\pi}^2)$.
If $C_{\mu\nu}^{V,A}(q,0)$ are represented through dispersion relations
by contributions of the physical states in $V$ and $A$ channels
($\rho$, $a_1$, $\pi$, etc), then according to Eq.(\ref{mix}) the poles
which are in the r.h.s. of Eq.(\ref{mix}), i.e. at $T=0$, appear at the
same positions in the l.h.s. Therefore, in order $T^2$ the poles
corresponding to $\rho$, $a_1$ or $\pi$ do not move\footnote{In Ref.\cite{dei}
it was not stressed that the poles do not move. Instead, the l.h.s.
of Eq.(\ref{mix}) was represented by {\em one} effective pole
($\rho$ in $V$-channel and $a_1$ in $A$-channel). Such a representation is
approximate, has no deep sense and corresponds to the description of the
current correlators at finite temperature by a one-pole contribution usually
used in lattice calculations of hadron masses. The fact that $\rho$ and $a-1$
poles do not move in order $T^2$ was overlooked in a recent paper\cite{hkl}.
For this reason the results obtained in Ref.\cite{hkl} are not reliable.}.
An important consequence of Eq.(\ref{mix}) is that at $T\neq 0$ in the
vector (transverse) channel apart from the poles corresponding to vector
particles, there arise poles corresponding to axial particles and visa verse,
i.e. a sort of parity mixing phenomenon occurs. The manifestation of this
phenomenon is in complete accord with the general considerations presented
above: the appearence of an $a_1$ pole in the vector channel corresponds to
singularities in the $s$-channel. In the same way a pion pole appears in the
longitudinal part of the vector channel.

The statement that the poles do not move in order $T^2$ is very general:
it is based only on PCAC and current algebra and can be immediately
extended to any other current correlators. (The result that the nucleon pole
does not move in order $T^2$ was obtained in the chiral perturbation theory
in Ref.\cite{ls} and by considering a current correlator in Ref.\cite{e2}).
The only interesting physical phenomenon  which occurs
in this order is the parity mixing, i.e. the appearence
of states with opposite parity in the given channel and, in some cases,
also an isospin mixing. The latter arises in baryon current correlators
where, for example, in the current with the quantum numbers of $\Lambda$ there
appears a $\Sigma$ pole, and in the nucleon channel there appear poles
corresponding to baryon resonances with $J^{P}=\frac{1}{2} ^{\pm}$ and
$T=\frac{3}{2} ,\;\frac{1}{2}$.

In the next order $O(T^4)$ such a simple picture where the current
correlator at finite temperature is represented by the superposition of
$T=0$ correlators does not take place. Interpreted in terms of temperature
dependent poles, it would mean that masses are shifted in this order.
But as explained above such an interpretation can be ambiguous.
We plan to discuss this problem in a future publication.

One of the authors (V.L.E.) would like to thank Joe Kapusta and Larry
McLerran, for the warm
hospitality during his stay at the Nuclear Theory Group and
Theoretical Physics Institute, the University of Minnesota.
This work was supported in part by the  Department of Energy under contract
DE-FG02-87ER40328.

\end{document}